\documentclass[conference]{IEEEtran}
\IEEEoverridecommandlockouts
\usepackage[utf8]{inputenc}
\usepackage{cite}
\usepackage{amsthm}    
\usepackage{amsmath}
\theoremstyle{plain}             

\usepackage{algorithm}
\usepackage{algpseudocode}
\usepackage{multirow} 
\usepackage{makecell}
\algrenewcommand\alglinenumber[1]{\scriptsize #1:}
\usepackage{graphicx}
\usepackage{textcomp}
\usepackage{newunicodechar}
\newunicodechar{↔}{\leftrightarrow}
\usepackage{xcolor}
\usepackage{booktabs}
\usepackage{caption}
\usepackage{url}
\usepackage{tikz}

\usepackage{listings}
\def\BibTeX{{\rm B\kern-.05em{\sc i\kern-.025em b}\kern-.08em
    T\kern-.1667em\lower.7ex\hbox{E}\kern-.125emX}}
\usepackage{listings}
\usepackage{xcolor}

\usepackage{xspace}

\newcommand{\approachCPUBaseline}{CPU Baseline\xspace}
\newcommand{\approachPIMBaseline}{Baseline PIM R-tree\xspace}
\newcommand{\approachBroadcastPIM}{Broadcast PIM R-tree\xspace}

\def\platformMill{\textsc{Mill}\xspace}
\def\platformPIM{\textsc{UPMEM-PIM}\xspace}

\def\MillCPUSeq{\textsc{CPU}-seq$_{\platformMill}$\xspace}
\def\MillCPUPar{\textsc{CPU}-par$_{\platformMill}$\xspace}

\def\PIMHost{\textsc{Host-Seq}$_{\platformPIM}$\xspace}
\def\PIMKernel{\textsc{Kernel}$_{\platformPIM}$\xspace}
\def\PIMEtwoE{\textsc{E2E}$_{\platformPIM}$\xspace}

\usepackage{todonotes}

\lstdefinestyle{ieee}{
  backgroundcolor=\color{gray!8},
  basicstyle=\ttfamily\footnotesize,
  numbers=left,
  numberstyle=\tiny\color{gray!70},
  stepnumber=1,
  numbersep=6pt,
  frame=single,
  rulecolor=\color{gray!50},
  frameround=tttt,
  breaklines=true,
  breakatwhitespace=true,
  tabsize=2,
  showstringspaces=false,
  columns=fullflexible,
  keepspaces=true,
  upquote=true
}

\begin{document}

\title{Parallel R-tree--based Spatial Query Processing on a Commercial Processing-in-Memory System\\
}


\author{
\IEEEauthorblockN{
Tasmia Jannat$^{1}$,
Michael Gowanlock$^{2}$,
Satish Puri$^{1}$
}

\IEEEauthorblockA{$^{1}$Department of Computer Science\\
Missouri University of Science and Technology, Rolla, MO, USA}

\IEEEauthorblockA{$^{2}$School of Informatics, Computing, and Cyber Systems\\
Northern Arizona University, Flagstaff, AZ, USA}

\IEEEauthorblockA{
tjv7w@mst.edu, michael.gowanlock@nau.edu, satish.puri@mst.edu
}
}

\maketitle

\begin{abstract}
The growing volume of data in scientific domains has made spatial query processing increasingly challenging due to high data transfer costs across the memory hierarchy and limited memory bandwidth. To address these bottlenecks and reduce the energy consumed on data movement, this work explores Processing-in-Memory (PIM) systems by executing range queries directly inside memory chips. Unlike prior PIM studies centered on linear scans or hash-based queries, this work is the first to map R-tree range queries onto commercial PIM hardware. The proposed broadcast-based method constructs the R-tree bottom-up on the CPU, broadcasts top levels to UPMEM DPUs (DRAM Processing Units) for global filtering, and distributes lower levels for parallel batched queries in a CPU–DPU system. We evaluate our approach on two real spatial datasets Sports (999K rectangles) and Lakes (8.4M rectangles) and assess scalability using a synthetic dataset with up to 16M rectangles and 3.9M queries, on a commercial UPMEM PIM system with up to 2,540 DPUs. Across all datasets, broadcast-based execution consistently outperforms subtree partitioning by preventing communication from dominating execution. On the Lakes dataset, strong scaling from 512 to 2,540 DPUs reduces kernel time from 64.9 s to 17.6 s, yielding up to 3.66× kernel and 2.70× end-to-end speedup relative to the CPU R-tree search on the same system. The PIM kernel also consumes approximately 3.4× less energy than the corresponding CPU search (e.g., 59.6 kJ vs. 167.0 kJ on Lakes), demonstrating scalable and energy-efficient hierarchical spatial range queries.
\end{abstract}

\begin{IEEEkeywords}
Range Search, Processing-in-Memory (PIM), High-Performance Computing (HPC), Spatial Query Processing, R-tree Structure
\end{IEEEkeywords}

\section{Introduction}
HPC systems face significant challenges in handling the exponential increase in data, especially in applications such as Geographic Information Systems (GIS) and spatial databases, which require efficient spatial query processing. Traditional computing architectures struggle because they face the memory wall problem \cite{wulf1995hitting}, the widening gap between processor speed and memory access, which hinders performance for data intensive tasks. This has led to interest in Processing-in-Memory (PIM) architectures that bring computation closer to the data, reducing bottlenecks, and improving both speed and energy efficiency. 
This limitation is especially evident in spatial query processing. Many scientific and geospatial workloads issue large numbers of range queries over rectangles and polygons. While R-tree indexing reduces the number of candidate objects examined, the cost of accessing index nodes and data records from memory frequently dominates both performance and energy consumption. PIM offers a direct solution by executing  computations near data. Among available platforms, UPMEM DPUs \cite{upmem2018whitepaper, gomez2022benchmarking} are notable for scale, programmability, and availability in server configurations, which makes them attractive for experiment driven systems work. 

Most prior PIM evaluations focus on scans and hash based operators \cite{baumstark2023accelerating},\cite{lim2023design}, \cite{kang2023pim}. In contrast, little is known about how a hierarchical spatial index such as the R-tree behaves on real PIM hardware, especially when the index is built on the host CPU and traversed across thousands of near memory cores with explicit host to device transfers and small local memories. 

This paper addresses this gap by presenting the first design and implementation of R-tree range query on a commercial PIM platform. We introduce a broadcast-based execution strategy that serializes and broadcasts compact upper-level R-tree headers to all DPUs for global pruning, while partitioning leaf nodes across DPUs for parallel processing. Queries are broadcast in batches, and each DPU applies upper-level filtering using the locally available R-tree headers before scanning its assigned leaf nodes, thereby preserving hierarchical pruning while minimizing communication and redundant data movement.

We evaluate our design on real and synthetic spatial datasets using a commercial UPMEM PIM system with thousands of DPUs. The results show that the proposed broadcast-based execution consistently outperforms subtree-based partitioning by preventing communication from dominating execution. Compared to CPU-based spatial search on the same system, our approach achieves substantial performance improvements and significantly lower energy consumption during the query phase. These findings indicate that hierarchical spatial query processing can be executed efficiently on PIM architectures when index layout and communications are carefully aligned with hardware constraints. We make the following contributions:

\begin{enumerate}
    \item \textbf{A heterogeneous CPU–PIM design for hierarchical spatial query processing.}
    We develop a CPU–DPU execution pipeline that constructs and serializes an R-tree on the CPU, broadcasts the upper levels to all DPUs, and distributes the leaf level in contiguous breadth-first partitions across ranks.
    
  \item \textbf{First R-tree implementation on commercial Processing-In-Memory architecture.}
    To our knowledge, this is the first study to map R-tree traversal onto a real PIM platform. The design broadcasts the top-level tree structure to all DPUs and distributes leaf partitions across DPUs to expose parallelism.

  
  \item \textbf{An empirical study of performance and energy on real workloads.}
We evaluate our design on two real spatial datasets (Sports and Lakes) and a large synthetic dataset with up to 16M rectangles and 3.99M queries, using up to 2,540 DPUs. Our study analyzes kernel-level and end-to-end performance, demonstrating up to 3.6× kernel speedup and 2.7× end-to-end speedup over the CPU search phase on the same system, along with up to 3.4× lower energy consumption for the PIM kernel.
\end{enumerate}

Overall, this work shows that hierarchical spatial query processing can be effectively supported on commercial PIM systems through hardware-aware index organization and communication design. By presenting the first implementation and evaluation of R-tree–based spatial query processing on a real PIM platform, the proposed approach highlights the potential of PIM architectures for accelerating irregular, data-intensive workloads beyond linear scans and hash-based operators.

\section{Background and Motivation}

\subsection{PIM System Architecture and Model}
\begin{figure}[ht]
  \centering
  \includegraphics[width=\linewidth]{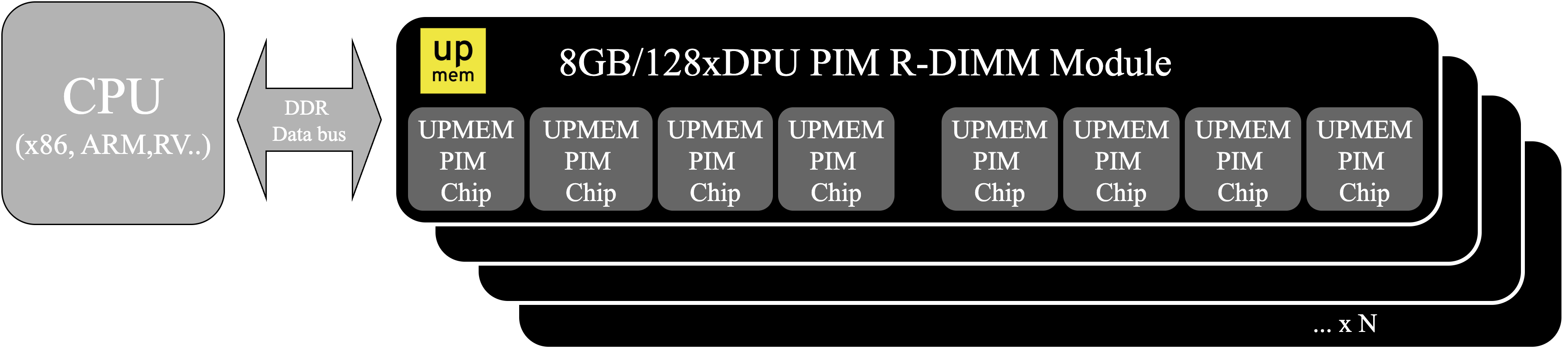}
\caption{Processing-in-Memory (PIM) System Organization \cite{devaux2019true}. DPU stands for DRAM Processing Unit.}
  \label{fig:system_organization}
\end{figure}

PIM architectures integrate simple processing elements directly within memory chips, enabling computation to be performed close to data and reducing costly data movement across the memory hierarchy \cite{patterson2002case}. Figure \ref{fig:system_organization} shows the UPMEM processing in memory system that has been used in our work. Each DIMM contains 16 PIM chips, and each chip integrates 8 DRAM Processing Units (DPUs) running at 400~MHz, resulting in 128 DPUs and 8~GB of DRAM capacity per DIMM. At server scale, a 20-DIMM configuration provides up to 2{,}560 DPUs and 160~GB of memory, enabling thousands of near-memory cores to operate in parallel. 

Each DPU integrates a lightweight processing core and a hierarchical memory system consisting of MRAM, WRAM, and instruction memory (IRAM). MRAM provides 64 MB of DRAM-based storage per DPU and is optimized for bulk accesses, making it suitable for large data structures such as serialized index nodes and leaf records. WRAM is a small (64 KB), low-latency memory shared among tasklets and is used for control metadata, query buffers, and intermediate results. IRAM stores DPU instructions and is managed by the runtime. DPUs support multiple lightweight hardware threads (tasklets); in the UPMEM PIM v1B platform used in this work, each DPU supports up to 16 tasklets, enabling intra-DPU parallelism under tight WRAM constraints.

DPUs do not support direct inter-DPU communication; all data exchange occurs via the host using explicit bulk transfers. Consequently, applications follow a bulk-synchronous parallel (BSP) execution model \cite{devaux2019true}, which favors communication-efficient designs that exploit local MRAM access and carefully manage limited WRAM.

\subsection{R-tree}

Spatial databases and GIS engines answer range-based queries, joins, and overlays on rectangles and polygons, which require indexing methods that can efficiently prune large portions of search space~\cite{puri2013mapreduce,spjoin16,rtreeOverlay,rtreeSpatialJoin22,hifire19}. Introduced by Guttman \cite{guttman1984r}, the R-tree is a height-balanced spatial index that organizes objects into a hierarchy of minimum bounding rectangles (MBRs). Leaf nodes store object MBRs, while internal nodes store MBRs that bound their children, enabling hierarchical pruning during query evaluation. Although node overlap can lead to multiple traversal paths, R-trees remain widely used in GIS, spatial databases, and scientific workloads due to their flexibility and effective support for spatial range queries. For static datasets, bulk-loading improves R-tree construction and query performance. One of the most widely adopted bulk-loading algorithms is the Sort-Tile-Recursive (STR) method proposed by Leutenegger et al. \cite{leutenegger1997str}. STR builds a packed R-tree bottom-up by sorting objects along one dimension, partitioning them into slices, and grouping them into leaf nodes with fixed fanout; higher levels are formed recursively from child MBRs. STR achieves high space utilization and reduced overlap compared to incremental insertion, improving query pruning. In this work, we adopt STR bulk-loading due to its low overlap, high space utilization, and predictable level-wise structure, which are well suited for parallel spatial query processing.

The parallel R-tree data structure has been well explored in the literature for multi-cores~\cite{kamel1992parallel}, GPUs~\cite{rtreeGPU2011,prasad2015gpu,kim2013parallel,you2013parallel,libRTS} and distributed memory environments~\cite{schnitzer1999master,rtree2000design,catFishRDMA,rtreeRDMA}. However, porting existing designs to the PIM system is not straightforward due to architectural differences between the traditional processor-centric and PIM paradigms.

\subsection{Prior Works: PIM-Friendly Indexes}

Most of the existing work on PIM focuses on data-parallel operators with regular access patterns, including table scans \cite{baumstark2023accelerating, gomez2022benchmarking}, selection and aggregation \cite{ahn2015pim}, and hash-based joins \cite{lim2023design}. Prior studies on commercial PIM platforms, particularly UPMEM, show that such operators benefit from high internal memory bandwidth and massive parallelism across DPUs \cite{lim2023design, gomez2022benchmarking}. Similar trends have also been reported in earlier near-data processing systems targeting analytics workloads \cite{ahn2015pim, balasubramonian2014near, seshadri2017ambit}.

Indexing support on PIM has received comparatively less attention. Existing PIM-friendly index designs primarily target key–value workloads and rely on simplified data structures such as hashing or skip lists to avoid deep pointer traversal and irregular control flow, as exemplified by PIM-Tree \cite{kang2023pim}. More recently, Kim et al. proposed OLTPim, an end-to-end OLTP DBMS for commercial UPMEM systems that offloads pointer-chasing components such as indexes and version chains to PIM, using hash-partitioned local B+trees and batched execution to amortize offload overheads \cite{kim2025no}. A recent work has also explored space-partitioning indexes for multidimensional points on PIM, such as PIM-zd-tree \cite{zhao2026pim}, which leverages space-filling curve ideas to linearize multidimensional data and reduce traversal complexity. 
In contrast, R-trees perform hierarchical pruning over MBRs (bounding rectangles) and require multi-level traversal with data-dependent branching, which makes them structurally different from the above PIM index designs. Moreover, R-tree range queries are output-sensitive: query performance depends not only on the input size and traversal depth, but also on the number of overlapping objects produced as output, which is not known a priori. It presents additional implementation challenges to handle variable size output. The proposed design handles a variety of output sizes across different queries and since the memory allocation is different in PIM, our implementation takes care of output-sensitive nature of the data structure.
To the best of our knowledge, we are not aware of any existing work that implements and evaluates R-tree range-query processing on commercial UPMEM PIM hardware; this work addresses this gap with a broadcast-based execution strategy that keeps host–DPU communication from dominating end-to-end execution.

\section{Methodology}

This section presents the three execution approaches considered in our study: a multi-threaded CPU implementation (\approachCPUBaseline), a subtree-based PIM design used as the baseline (\approachPIMBaseline), and the proposed broadcast-based PIM design (\approachBroadcastPIM).

\subsection{Multi-threaded CPU Baseline}
\label{sec:cpu_baseline}

We compare PIM-based execution against a multi-threaded CPU baseline that performs R-tree range queries entirely in host memory. To ensure a fair comparison, the CPU baseline uses the \emph{same R-tree structure} as the proposed broadcast-based PIM approach, constructed on the host with identical bulk-loading parameters. This ensures that performance differences arise from the execution model rather than from index structure.

R-tree construction is performed sequentially, as it is a one-time preprocessing cost, while query processing is parallelized across CPU threads. Each query traverses the R-tree using bounding-box filtering followed by exact rectangle
intersection tests, matching the semantics of the PIM kernel.

We implement a custom multi-threaded search engine using POSIX threads, as publicly available R-tree libraries do not provide scalable, thread-parallel query execution with explicit control over scheduling and traversal behavior. Parallel query processing uses dynamic, chunk-based scheduling to mitigate load imbalance caused by spatial skew and query selectivity. The R-tree is read-only during query execution, eliminating synchronization within the traversal. Algorithm~\ref{alg:cpu_parallel_search} outlines the parallel CPU query processing strategy.

\begin{algorithm}[t]
\caption{Parallel CPU R-tree Range Query Processing (CPU Baseline)}
\label{alg:cpu_parallel_search}
\begin{algorithmic}[1]
\Require R-tree root $R$, query set $Q[0 \ldots N-1]$, threads $T$, chunk size $C$
\Ensure Result array $results[0 \ldots N-1]$
\State Initialize shared atomic index $idx \gets 0$
\ForAll{threads $t = 0$ to $T-1$ \textbf{in parallel}}
    \While{true}
        \State $start \gets$ atomic\_fetch\_and\_add$(idx, C)$
        \If{$start \ge N$} \textbf{break} \EndIf
        \State $end \gets \min(start + C, N)$
        \For{$i = start$ to $end - 1$}
            \State $results[i] \gets$ \Call{SearchR-tree}{$R, Q[i]$}
        \EndFor
    \EndWhile
\EndFor
\State \Return $results$
\end{algorithmic}
\end{algorithm}


\subsection{Subtree-based \approachPIMBaseline}
\label{sec:subtree}

As a baseline, we implement a subtree-based PIM R-tree in which each DPU is assigned an independent R-tree subtree and processes queries locally. This design aligns with PIM systems, where DPUs operate on DPU-local MRAM address spaces without inter-DPU communication. We use this baseline to quantify the cost of subtree partitioning and host–DPU data movement.

\paragraph{Fanout-Constrained R-tree Construction}
To enforce a one-to-one mapping between subtrees and DPUs, we construct the R-tree on the host using a custom top-down bulk-loading strategy that explicitly constrains the root fanout to the number of DPUs. Traditional R-tree constructions are not suitable for this as Guttman’s insertion-based R-tree yields data-dependent fanout, while STR bulk loading builds the tree bottom-up without controlling the number of top-level subtrees. Algorithm~\ref{alg:subtree_partition} outlines the construction, which recursively partitions rectangles using x- and y-center ordering to form spatially coherent groups while capping the root fanout. Although not a classical STR construction, this approach adopts STR-style spatial ordering to improve locality, reduce overlap, and balance subtree sizes.

\begin{algorithm}[t]
\caption{Fanout-constrained R-tree creation in PIM Baseline}
\label{alg:subtree_partition}
\begin{algorithmic}[1]
\Require Rectangles $R$, number of DPUs $P$, leaf capacity $B$
\Function{Build}{ Rectangles $R$ }
    \If{$|R| \le B$} \Return leaf node over $R$ \EndIf
    \State $k \gets \min(P,\lceil |R|/B\rceil)$
    \Comment{target number of children}
    \State Sort $R$ by x-center and split into $\lceil\sqrt{k}\rceil$ slabs
    \ForAll{slabs}
        \State Sort by y-center and partition into groups
        \State Recursively build child nodes for each group
    \EndFor
    \State \Return internal node with $\le k$ children
\EndFunction
\State Build root $T \gets \textsc{Build}(R)$ and assign its children as one subtree per DPU
\end{algorithmic}
\end{algorithm}

\paragraph{Subtree Serialization and Execution}

\begin{figure}[h]
  \centering
  \includegraphics[width=0.88\linewidth]{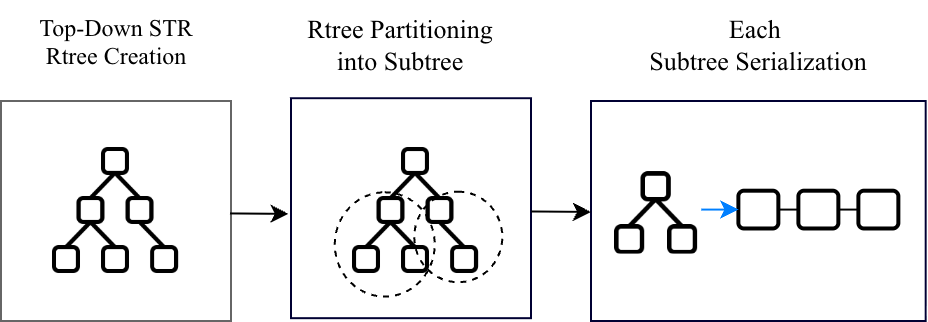}
\caption{Three-step subtree-based PIM R-tree (PIM baseline) preparation:
(1) top-down STR-inspired R-tree construction on the CPU,
(2) partitioning at the root into level 1 subtrees,
and (3) serialization of each subtree for DPU transfer.
}
\label{fig:subtree_serialization}

\end{figure}

Figure~\ref{fig:subtree_serialization} illustrates the baseline workflow. After construction, the R-tree is partitioned at the root, and each level-1 subtree is serialized into a contiguous representation and transferred to its assigned DPU. Each DPU stores its subtree in MRAM and evaluates all queries locally using multiple tasklets. 

In this \approachPIMBaseline, query rectangles are broadcast to all DPUs in batches. Each DPU evaluates the complete query set independently against its locally stored subtree using \(T\) tasklets. Queries are statically partitioned across tasklets to eliminate synchronization overhead and avoid write contention during result accumulation. For each query, the DPU performs a recursive R-tree traversal starting from the root of its assigned subtree. At internal nodes, MBR intersection tests are applied to prune non-overlapping branches. At leaf nodes, all contained rectangles are examined for overlap with the query region, and the number of intersecting rectangles is counted. Each DPU produces a per-query overlap count corresponding to its subtree, and the host aggregates the partial counts returned by all DPUs to obtain the final query result.

This design exposes parallelism across DPUs, but requires transferring a distinct serialized subtree to each DPU, leading to substantial host to DPU communication overhead. We use this subtree-based design as a PIM baseline to quantify the cost of per-DPU subtree transfers and motivate the broadcast-based execution strategy introduced next.

\subsection{Proposed \approachBroadcastPIM}
\begin{figure}[h]
  \centering
  \includegraphics[width=\linewidth]{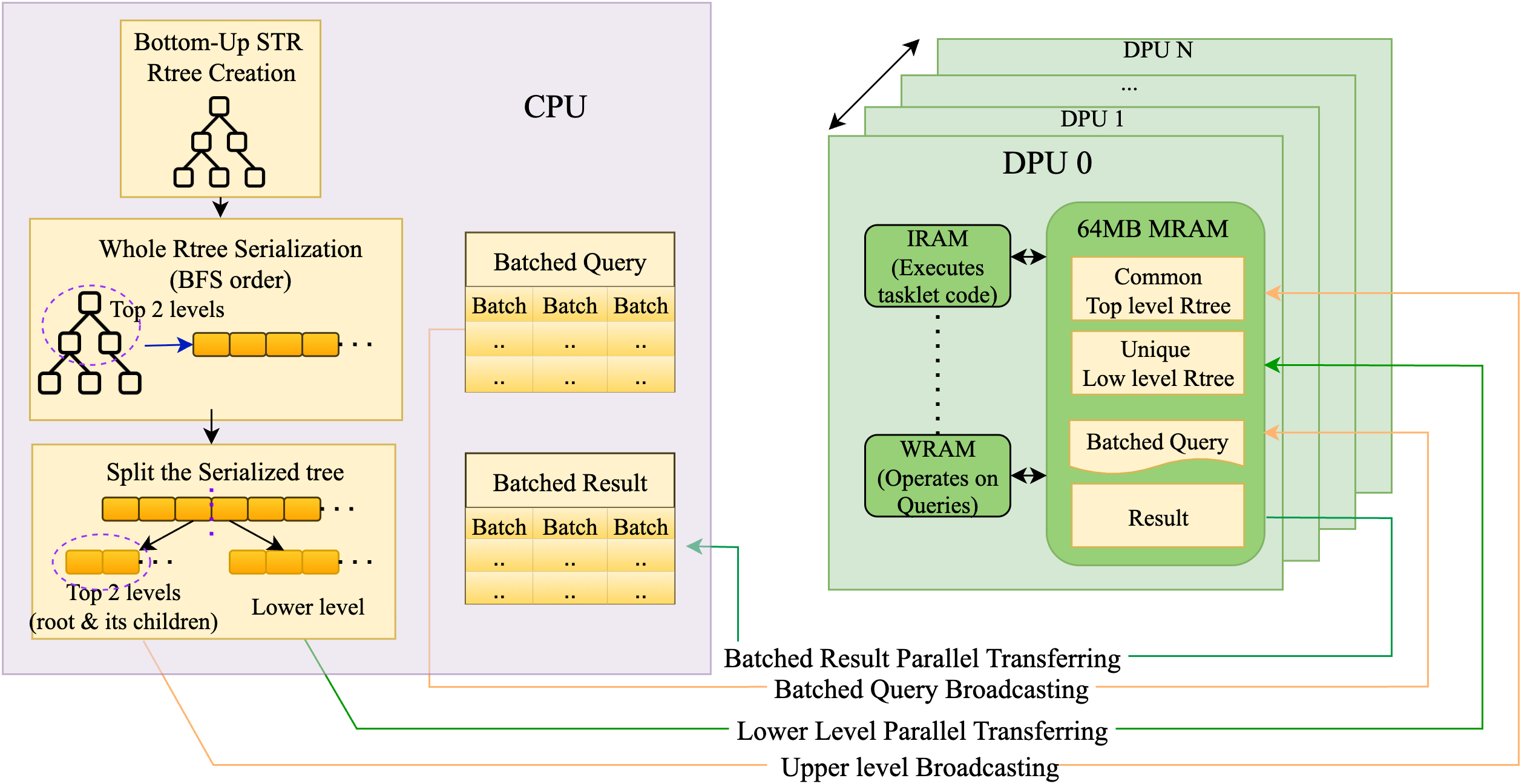}

\caption{Workflow for \approachBroadcastPIM range query processing on a heterogeneous CPU–PIM system.}
  \label{fig:broadcasting}
\end{figure}

We implement a heterogeneous CPU to PIM workflow in which the host constructs an R-tree, broadcasts a compact prefix of the index to all DPUs, partitions the leaf level across devices, and evaluates batched queries in parallel with device-side pruning and MRAM-based leaf scans. In this heterogeneous system, the CPU orchestrates data distribution and query scheduling, and PIM hardware executes searches in parallel. Figure \ref{fig:broadcasting} shows the workflow.

\subsubsection{CPU-side R-tree construction}

\begin{figure}[h]
  \centering
  \includegraphics[width=0.98\linewidth]{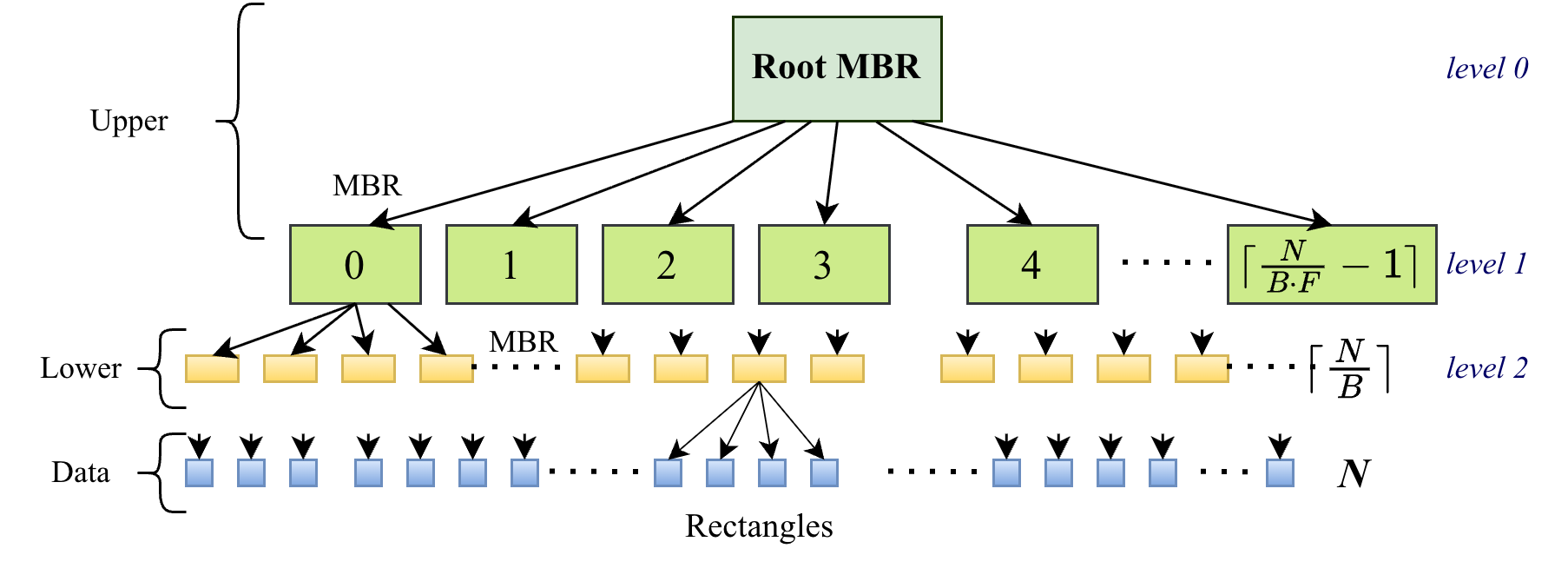}
\caption{Three-level STR R-tree layout for a dataset of \(N\) rectangles. Each leaf node stores up to \(B\) rectangles, and internal nodes at level~1 have fanout \(F\), where \(F\) equals the number of DPUs. The tree contains \(\lceil N/B \rceil\) leaf nodes and \(\lceil N/(B\,F) \rceil\) level 1 internal nodes.}

\label{fig:str-3level-layout}

\end{figure}

We construct the R-tree on the CPU using the standard bottom-up STR bulk-loading procedure, with fixed leaf and internal node capacities; the choice of parameters bounds the tree height but does not alter the STR grouping strategy. Let $N$ denote the number of input rectangles. At the leaf level, rectangles are sorted by the midpoint of their horizontal extent
($x_{\text{center}}=(x_{\min}+x_{\max})/2$) and partitioned into $\lceil\sqrt{N/B}\rceil$ contiguous slices, where $B$ is the leaf capacity (\texttt{BUNDLEFACTOR}). Within each slice, rectangles are sorted by the midpoint of their vertical extent
($y_{\text{center}} = (y_{\min} + y_{\max})/2$) and packed into leaf nodes of fixed capacity. Internal nodes are constructed recursively using the same STR grouping strategy: child nodes are treated as objects, sorted by the midpoint of their MBR extents, tiled into slices, and packed into parent nodes with fanout bounded by \texttt{FANOUT}. This process continues until a single root is formed. We select \texttt{BUNDLEFACTOR} and \texttt{FANOUT} such that the resulting R-tree has exactly three levels, with all leaves at level~2, as illustrated in Figure~\ref{fig:str-3level-layout}. This three-level design keeps the broadcast prefix (root + level-1) within DPU WRAM. Deeper trees would require broadcasting additional internal levels whose size can approach \texttt{FANOUT}, making WRAM a practical constraint.

\subsubsection{BFS serialization on the host}
To enable DPU execution, the entire R-tree is serialized into a flat array of nodes, replacing pointers with array indices. This is necessary because UPMEM DPUs do not support dynamic allocation or host pointers; all device memory must be statically allocated and index-addressable. 

\begin{lstlisting}[language=C, caption={SerializedNode (SN) structure used for breadth-first serialization.}, label={lst:serialized-node}]
typedef struct SN {
    int isLeaf;
    int count;
    MBR mbr;
    int  children[MAX_CHILDREN];
    Rect rects[MAX_RECTS];
} SN;
\end{lstlisting}

Here, \texttt{MAX\_CHILDREN} equals the internal node fanout (\texttt{FANOUT}), and \texttt{MAX\_RECTS} equals the leaf capacity (\texttt{BUNDLEFACTOR}). Serialization produces a contiguous array \(\texttt{SN}[0..K-1]\), where \(K\) is the total number of serialized nodes, laid out in breadth-first order. The root is stored at index~0, followed by all level 1 internal nodes and then all level 2 (leaf) nodes. With this layout, the beginning of the leaf level can be computed directly as
\(1 + \texttt{SN}[0].\texttt{count}\), where \texttt{SN}[i].\texttt{count} denotes the number of children for internal node \(i\) (or the number of rectangles for a leaf node).

Serialization is performed in a single breadth-first pass: the total node count \(K\) is computed in advance, memory is allocated once, and each node is written exactly once, yielding \(O(K)\) time complexity. Although depth-first layouts are common for traversal, our objective is efficient data placement and transfer. A breadth-first layout forms a compact prefix containing the shared upper levels, which can be broadcast once to all DPUs, while leaf nodes remain contiguous for coalesced per-DPU transfers. On UPMEM systems, broadcast transfers scale better with DPU count than repeated per-DPU copies; we exploit this property by broadcasting the shared upper levels only once~\cite{cai2024pimpam}.

\subsubsection{R-tree transfer to DPUs}

R-tree transfer is carried out in two stages: (1) a one-time broadcast of the shared upper levels of the R-tree and (2) a parallel distribution of the remaining leaf nodes, which are local to individual DPUs.

\paragraph{Broadcast of the upper levels}
We broadcast the shared upper two levels once to all DPUs. Owing to breadth-first serialization, the root and all level~1 internal nodes form a contiguous prefix \(\texttt{SN}[0..c]\), where \(c\) denotes the number of children of the root node. Since upper-level pruning requires only node metadata (\texttt{isLeaf}, \texttt{count}, and \texttt{mbr}), the host broadcasts a compact header representation instead of full nodes \(\texttt{SN}\). This reduces host to DPU communication volume and initialization overhead while preserving correctness, as upper-level filtering relies solely on MBR–query intersection tests.

\paragraph{Parallel transfer of leaf nodes}
The level~2 nodes (leaf) of the R-tree, are partitioned into contiguous slices and distributed across DPUs. Level~2 begins at index \(1 + \texttt{SN}[0].\texttt{count}\), and the slices are sized to balance the number of leaf nodes per DPU. Each slice is transferred using parallel, coalesced host to DPU transfers and stored in the DPU’s local MRAM. This data placement strategy balances workload across DPUs, eliminates inter-DPU communication, and enables fully independent parallel query execution.

\subsubsection{Query broadcast and batching}

Query rectangles are broadcast from the host to all DPUs in batches so that each query is evaluated against all leaf partitions. Broadcasting is required for correctness because each DPU stores only a disjoint subset of the leaf nodes, and a query may intersect leaves residing on multiple DPUs; partial results are therefore aggregated on the host. For each batch, the host broadcasts the batch size and the corresponding query rectangles, amortizing communication overhead and enabling reuse of the broadcasted upper-level metadata.

\subsubsection{Parallel batch execution on DPUs}

After receiving a query batch, each DPU processes the queries independently using multiple tasklets. Let \(F\) denote the number of DPUs and \(T\) the number of tasklets per DPU. Each DPU \(d \in \{0,\dots,F-1\}\) stores a contiguous slice of the serialized R-tree leaf level in its local MRAM. We denote this slice by \(\mathcal{L}_d \subset \texttt{SN}\), where \(\mathcal{L}_d\) consists exclusively of level~2 (leaf) nodes assigned to DPU \(d\). At kernel launch, the broadcast upper-level metadata (root and level~1 nodes) is read from MRAM into WRAM using \texttt{mram\_read}. This metadata is shared by all tasklets on the DPU. A barrier ensures that the WRAM-resident upper-level data is available before query processing begins. Queries in the current batch are statically partitioned across tasklets to ensure load balance and to avoid synchronization or write contention. Each tasklet processes a disjoint subset of queries and operates independently.

\paragraph{Two-phase query evaluation}
For each assigned query, a tasklet performs a two-phase search.
In \emph{Phase~1 (upper-level filtering)}, as shown in Algorithm \ref{alg:dpu_parallel} and Figure \ref{fig:dpu_level1_mapping}, the tasklet evaluates the query against a small, constant number (at most four) of level~1 MBRs. This bounded neighborhood ensures correctness near region boundaries. These MBRs are initially stored in MRAM and fetched once into WRAM. Because the upper levels contain only a small number of nodes, they fit easily within the DPU’s WRAM capacity and can be reused across all queries. The candidate level~1 nodes are determined by the DPU index and the root fanout, reducing the upper-level filtering cost from \(O(c)\) to \(O(1)\), where \(c\) denotes the number of children of the root. If no overlap is detected at this stage, the query is discarded locally.
 
In \emph{Phase~2 (local leaf scan)}, each tasklet sequentially scans the leaf nodes in \(\mathcal{L}_d\), where \(\mathcal{L}_d\) denotes the subset of level~2 (leaf) nodes assigned to DPU \(d\). The tasklet reads leaf data directly from MRAM and applies rectangle–query overlap tests. For each overlapping rectangle, a local counter is incremented. Because the leaf data are read-only during this phase, the scan is performed without buffering rectangles in WRAM. Per-query overlap counts are accumulated in WRAM and written back to MRAM upon completion. Tasklets execute independently without inter-tasklet communication.

Kernel execution time on a DPU is defined as the maximum execution time across its tasklets, since tasklets execute concurrently and overall completion is determined by the slowest tasklet. This per-DPU kernel time is used as the batch execution time in our analysis.

\begin{algorithm}[t]
\caption{Parallel Batch Execution on each DPU}
\label{alg:dpu_parallel}
\begin{algorithmic}[1]
\Require
Upper-level headers in MRAM \(\texttt{Hdr}[0..c]\), $c$ children,
local leaf slice \(\mathcal{L}_d\),
query batch \(Q_b\),
tasklets \(T\)
\Ensure
Per-query overlap counts stored in variable \(\texttt{DPU\_OVERLAP\_COUNT}\)

\State \textbf{Tasklet 0:} Read \(\texttt{Hdr}[0..c]\) from MRAM into WRAM
\State Barrier synchronization
\State Partition queries in \(Q_b\) evenly across tasklets

\For{each tasklet \(t\) \textbf{in parallel}}
    \For{each assigned query \(q\)}
        \State \textbf{Phase 1:} Test overlap with at most four level~1  MBRs (WRAM)
        \If{overlap detected}
            \State \textbf{Phase 2:} Scan leaf nodes in \(\mathcal{L}_d\) (MRAM) and count overlaps
        \Else
            \State Set overlap count to 0
        \EndIf
    \EndFor
\EndFor

\State Reduce per-tasklet cycle counts to obtain per-DPU kernel time
\end{algorithmic}
\end{algorithm}

\begin{figure}[h]
  \centering
  \includegraphics[width=\linewidth]{image/upper_level_filtering.png}
\caption{DPU-index--guided upper-level filtering in \approachBroadcastPIM.
DPUs are mapped to level~1 regions based on their index. Each DPU evaluates overlap against only a small neighborhood of adjacent level~1 MBRs (highlighted), reducing upper-level filtering from \(O(c)\) to \(O(1)\) per query.}
\label{fig:dpu_level1_mapping}

\end{figure}

\section{Implementation}
\label{sec:implementation}

We implement both the \approachPIMBaseline and the proposed \approachBroadcastPIM on a commercial UPMEM system. The host-side code is written in C and uses the UPMEM SDK to manage DPU allocation, data movement, and kernel execution, while DPU kernels are implemented in C and execute directly on the DPUs. The implementation follows a host–device execution model in which the CPU performs index construction and orchestration, and DPUs execute parallel spatial query processing within memory.

\subsection{Host-side implementation}

The host is responsible for R-tree construction, serialization, query preparation, and result aggregation. For both methods, the R-tree is constructed on the CPU using bulk-loading and serialized into a contiguous, index-addressable representation suitable for DPU execution. For the subtree-based baseline, the host partitions the R-tree at the root into disjoint subtrees, serializes each subtree independently, and transfers one serialized subtree to each DPU. Query rectangles are broadcast in batches, and each DPU processes all queries against its local subtree.

For the \approachBroadcastPIM method, the host serializes the entire R-tree in breadth-first order and separates it into two parts: (i) a shared prefix consisting of the root and level~1 internal nodes, and (ii) the level~2 leaf nodes. The shared prefix is broadcast once to all DPUs using \texttt{dpu\_broadcast\_to}, while the leaf nodes are partitioned into contiguous slices and transferred to DPUs using parallel, coalesced host-to-DPU transfers. Query rectangles are broadcast to all DPUs in batches. After kernel execution, per-query overlap counts are retrieved from DPUs and aggregated on the host.

\subsection{DPU-side implementation}

Each DPU stores its assigned tree data and query batches in MRAM. Query processing is parallelized across multiple tasklets within each DPU. Upon kernel launch, a designated tasklet initializes shared state, fetches any required upper-level metadata from MRAM into WRAM, and configures the hardware performance counter. A barrier ensures that all tasklets begin query processing with a consistent view of shared data. In the subtree-based baseline, each DPU traverses its local serialized subtree recursively for every query, applying MBR pruning at internal nodes and scanning leaf rectangles to count overlaps.

In the \approachBroadcastPIM, each DPU processes only its assigned slice of leaf nodes. For each query, tasklets perform a two-phase evaluation: a lightweight upper-level filtering phase using WRAM-resident metadata, followed by a local leaf scan directly from MRAM when filtering succeeds. Queries are statically partitioned across tasklets to ensure load balance and to avoid synchronization or write contention. Tasklets operate independently and do not communicate during execution.

\subsection{Performance measurement}

Each tasklet records its execution cycles using the DPU hardware performance counter and stores the value in the \texttt{DPU\_PERF} structure. For each DPU, the kernel execution time is defined as the maximum cycle count across its tasklets, reflecting the completion time of parallel execution. Host–DPU transfer times and end-to-end execution times are measured using host-side timers and are reported separately in the evaluation.

\paragraph{Energy measurement}
Energy consumption is measured using an external PN150 power meter connected to the UPMEM server power supply. Power is sampled during execution of the CPU-only and DPU-accelerated phases and combined with the corresponding execution time to compute energy consumption. These measurements are synchronized with kernel execution and host–DPU transfers. Detailed power states, datasets, and energy-efficiency results are reported in Section~\ref{sec:energy}.

\section{Experimental Evaluation}
This section evaluates the performance, scalability, and energy efficiency of our \approachBroadcastPIM and contrasts it with both a multi-threaded \approachCPUBaseline and a \approachPIMBaseline. Our evaluation is structured to examine three aspects of PIM-based spatial query processing: $(i)$ when PIM execution outperforms CPU-based approaches, $(ii)$ how broadcast-based execution compares to subtree partitioning (\approachPIMBaseline) in terms of communication efficiency and scalability, and, $(iii)$ how memory access patterns
influence performance and energy behavior on commercial PIM hardware. We first describe the experimental setup, datasets, and timing methodology, and then present detailed performance and energy results across a range of workloads.

\subsection{Experimental Setup}


We evaluate our experiments on two systems: \platformMill{} and \platformPIM{}. CPU baselines are measured on a shared university-operated HPC cluster named \platformMill{} with dual-socket Intel Xeon Gold 6248 processors (20 cores per socket, 40 cores total) running at a base frequency of 2.50\,GHz. CPU experiments use 8 threads for parallel execution and a single thread for sequential baselines. The thread count is chosen to represent a moderate level of parallelism and to avoid memory bandwidth saturation. All CPU runs use the same R-tree structure and query semantics as the PIM implementation to ensure comparability. PIM experiments are conducted on a commercial UPMEM system with up to 2{,}540 DPUs running at 400 MHz with 11 tasklets (out of 16) per DPU, where the host CPU shares memory channels with PIM DIMMs. Although the platform nominally supports 2{,}560 DPUs (20 DIMMs), allocations beyond 2{,}540 DPUs fail with a dpu allocation error, so we use 2{,}540 DPUs as the maximum stable configuration. We use 11 tasklets since performance saturates beyond 11. DPUs execute independently without inter-DPU communication. Queries are processed in batches of up to 10{,}000, and partial results are aggregated on the host.

We note that CPU performance on \platformMill{} represents an optimistic CPU-only baseline without PIM-induced memory bandwidth contention, while CPU execution on UPMEM reflects a system-level comparison where the host shares memory channels with PIM DIMMs. These two provide complementary perspectives for evaluating PIM performance.

\paragraph{Datasets}

We evaluate our approach using both real-world and synthetic spatial datasets that span a wide range of sizes, spatial distributions, and memory behaviors. The real-world datasets are drawn from UCR-STAR~\cite{ghosh2019ucr}, while the synthetic dataset is generated using SPIDER~\cite{spiderweb2021}. All datasets consist of rectangles represented as $(x_{\min}, y_{\min}, x_{\max}, y_{\max})$. To support execution on UPMEM PIM hardware, which does not efficiently support floating-point operations, all coordinates are converted to 32-bit integers using a fixed-precision scaling scheme. Together, these datasets allow us to study workloads that are cache-friendly (small datasets), moderately memory-intensive, and heavily memory-bound. For each dataset, we vary the number of range queries from approximately 1\% to 25\% of the whole dataset, enabling us to evaluate both low  and high query pressure regimes. Table~\ref{tab:dataset_summary} summarizes the datasets used in our experiments.
\begin{table}[t]
\centering
\footnotesize 
\caption{Summary of datasets and query workloads.} 
\label{tab:dataset_summary}
\begin{tabular}{|l|r|l|}
\hline
\textbf{Dataset} &
\makecell{\textbf{Dataset Size}\\(\#MBRs)} &
\makecell{\textbf{Query Sizes}\\(\#Queries)} \\
\hline
\multirow{2}{*}{\makecell[l]{Sports\\\textit{(UCR-STAR)}}}
 & \multirow{2}{*}{999K}
 & 9K (1\%), 49K (5\%) \\
 & & 99K (10\%), 249K (25\%) \\
\hline
\multirow{2}{*}{\makecell[l]{Lakes\\\textit{(UCR-STAR)}}}
 & \multirow{2}{*}{8.4M}
 & 84K (1\%), 420K (5\%) \\
 & & 841K (10\%), 2.1M (25\%) \\
\hline
\multirow{2}{*}{\makecell[l]{Synthetic\\\textit{(SPIDER)}}}
 & \multirow{2}{*}{16M}
 & 159K (1\%), 798K (5\%) \\
 & & 1.6M (10\%), 3.99M (25\%) \\
\hline
\end{tabular}
\end{table}


\paragraph{Timing Methodology}
We measure \MillCPUSeq{} and \MillCPUPar{} as end-to-end wall-clock execution time on \platformMill{}. These measurements include the complete CPU query-processing pipeline, covering query input handling, R-tree traversal, overlap testing, and result aggregation. \MillCPUSeq{} corresponds to single-threaded execution, while \MillCPUPar{} corresponds to multi-threaded execution using 8 CPU threads.

On \platformPIM{}, we separately measure kernel execution time and end-to-end execution time. \PIMKernel{} measures the DPU kernel execution time only and is obtained using the DPU hardware cycle counter, taking the maximum cycle count across tasklets to represent kernel completion. \PIMEtwoE{} measures the full end-to-end execution time on \platformPIM{}, including host to DPU data transfers, DPU kernel execution, and host-side result aggregation.


\subsection{Comparison with \approachCPUBaseline}\label{sec:cpu_baselines}
Host execution on a PIM-equipped system is constrained by reduced host memory bandwidth because part of the memory bandwidth is used by the PIM DIMMs\cite{friesel2023full}. To ensure a fair evaluation, we report two baselines. The first is \approachCPUBaseline measured on \platformMill, which uses only host DRAM and provides an optimistic reference without PIM-induced bandwidth contention. The second is measured on the \platformPIM{}, where the CPU shares memory channels with PIM DIMMs, enabling a direct system-level comparison with PIM end-to-end execution.

\begin{table}[t]
\centering
\footnotesize
\setlength{\tabcolsep}{3pt}
\renewcommand{\arraystretch}{0.95}
\caption{\approachBroadcastPIM execution on \platformPIM{} versus parallel \approachCPUBaseline on \platformMill{}. Cross-platform speedups report PIM kernel and end-to-end performance relative to CPU parallel execution on \platformMill{}.}
\label{tab:pim_vs_cpupar}
\begin{tabular}{l r r r r r r r}
\toprule
\multirow{2}{*}{\makecell{Data\\set}} &
\multirow{2}{*}{Query} &
\multicolumn{2}{c}{\platformMill{}} &
\multicolumn{2}{c}{\platformPIM{}} &
\multicolumn{2}{c}{\makecell{Cross-platform\\Speedup}} \\
\cmidrule(lr){3-4} \cmidrule(lr){5-6} \cmidrule(lr){7-8}
& &
\makecell{CPU\\Seq} &
\makecell{CPU\\Par} &
Kernel &
E2E &
\makecell{PIM\\Kernel} &
\makecell{PIM\\E2E} \\
\midrule

\multirow{4}{*}{Sports}
 & 1\%  & 0.42 & 0.45 & 0.30 & 1.10 & 1.50 & 0.41 \\
 & 5\%  & 2.16 & 0.45 & 1.50 & 2.77 & 0.30 & 0.16 \\
 & 10\% & 4.22 & 0.91 & 3.00 & 4.85 & 0.30 & 0.19 \\
 & 25\% & 10.42 & 1.80 & 7.52 & 11.07 & 0.24 & 0.16 \\
\midrule

\multirow{4}{*}{Lakes}
 & 1\%  & 15.32 & 2.59 & 3.61 & 6.00 & 0.72 & 0.43 \\
 & 5\%  & 74.38 & 10.98 & 17.57 & 23.85 & 0.62 & 0.46 \\
 & 10\% & 150.74 & 20.69 & 35.92 & 46.93 & 0.58 & 0.44 \\
 & 25\% & 361.71 & 49.80 & 87.73 & 112.90 & 0.57 & 0.44 \\
\midrule

\multirow{4}{*}{Synthetic}
 & 1\%  & 18.65 & 2.58 & 1.55 & 5.31 & \textbf{1.66} & 0.49 \\
 & 5\%  & 93.63 & 12.81 & 7.76 & 17.49 & \textbf{1.65} & 0.73 \\
 & 10\% & 187.07 & 25.49 & 15.54 & 32.78 & \textbf{1.64} & 0.78 \\
 & 25\% & 469.97 & 64.47 & 39.03 & 79.26 & \textbf{1.65} & 0.81 \\
\bottomrule
\end{tabular}
\end{table}

Table \ref{tab:pim_vs_cpupar}  compares the execution time and speedup of the proposed \approachBroadcastPIM against \approachCPUBaseline across three datasets and varying query selectivities. We report (i) sequential and parallel CPU execution on the \platformMill, (ii) DPU kernel time \PIMKernel{} and total DPU time, \PIMEtwoE{} on the \platformPIM{}, and (iii) PIM's speedups relative to the parallel \approachCPUBaseline. The results show \approachBroadcastPIM achieves up to 1.66× kernel speedup vs. \approachCPUBaseline on the Synthetic dataset, but does not outperform on the real datasets in this configuration. This behavior is consistent with prior PIM studies, where optimized multi-threaded CPU implementations often remain competitive or superior for complex data structures. 

\begin{figure}[h]
  \centering
  \includegraphics[width=\linewidth]{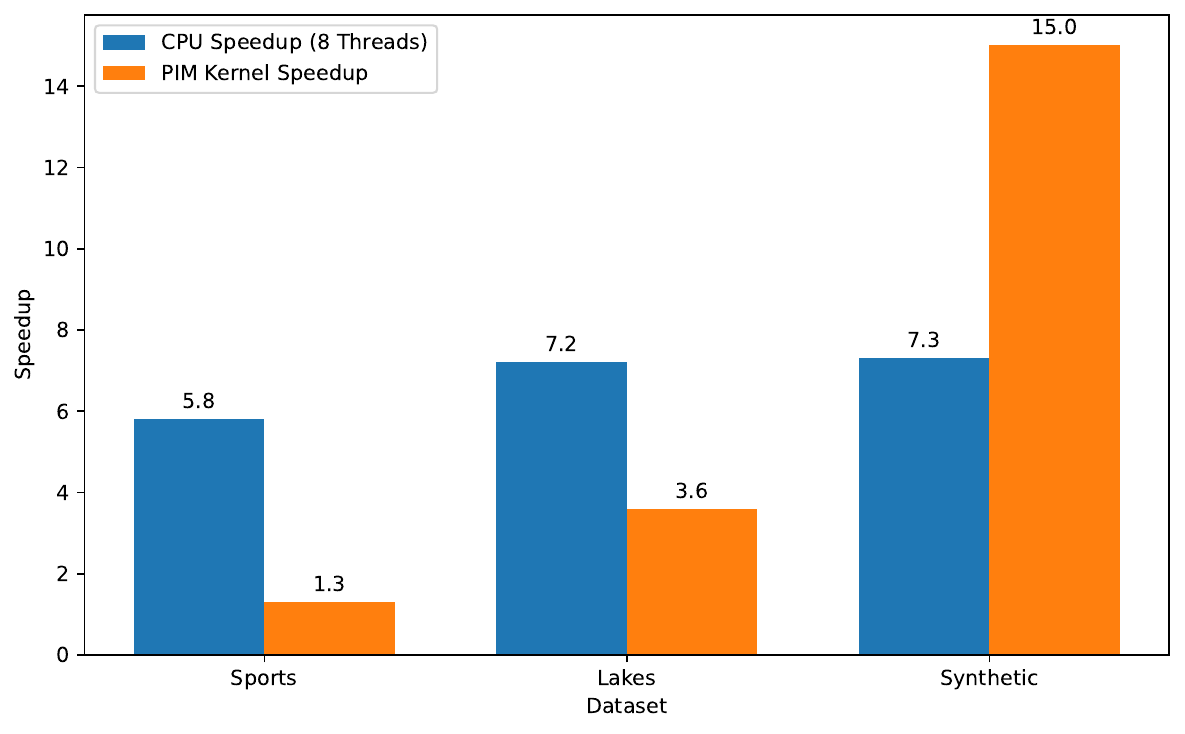}
\caption{Comparison of CPU-parallel speedup on \platformMill and PIM kernel speedup on \platformPIM{} across datasets. CPU speedup plateaus at approximately 6--7$\times$, while PIM kernel speedup increases substantially with dataset size. Here, CPU speedup is relative to CPU sequential execution on MILL, while PIM kernel speedup is relative to CPU sequential execution on UPMEM-PIM.} 
\label{fig:Cpu-dpu_kernel}

\end{figure}

Nevertheless, the observed kernel-level speedups on the Synthetic workload indicate that, when memory-level parallelism is effectively exploited and access patterns are well balanced, PIM execution can exceed CPU-parallel performance. These results highlight the potential of broadcast-based in-memory filtering and motivate further optimization of data placement, query batching, and communication overheads to close the gap for real-world workloads. While CPU parallelization on \platformMill{} achieves up to $7\times$ speedup over sequential execution, performance improvements diminish as query volume increases at the evaluated thread count. This behavior suggests that query processing becomes increasingly constrained by shared-memory bandwidth and cache contention rather than compute capacity. Figure~\ref{fig:Cpu-dpu_kernel} highlights the fundamental difference between CPU and PIM execution: CPU performance is limited by a shared-memory subsystem, whereas PIM execution distributes memory accesses across independent DPUs. As a result, PIM benefits from memory-intensive workloads, achieving kernel speedups of up to $15\times$ on the Synthetic dataset, where query processing is dominated by irregular memory accesses. Here, CPU speedup is computed relative to CPU sequential execution on \platformMill, and PIM kernel speedup is computed relative to sequential CPU phase execution on \platformPIM. 

\subsection{Comparison of \approachBroadcastPIM and \approachPIMBaseline Execution}
Table~\ref{tab:broadcast_vs_subtree}, which reports results measured entirely on the \platformPIM{}, shows that the \approachBroadcastPIM consistently outperforms the subtree-based baseline across all datasets and query sizes. While both approaches achieve comparable kernel speedups, the subtree-based design suffers from substantially higher end-to-end execution time due to excessive host to DPU communication. This effect is particularly pronounced for the Lakes dataset, where \approachBroadcastPIM execution achieves end-to-end speedups of up to $2.83\times$, compared to less than $0.5\times$ for the \approachPIMBaseline approach.

\begin{figure}[t]
  \centering
  \includegraphics[width=\columnwidth]{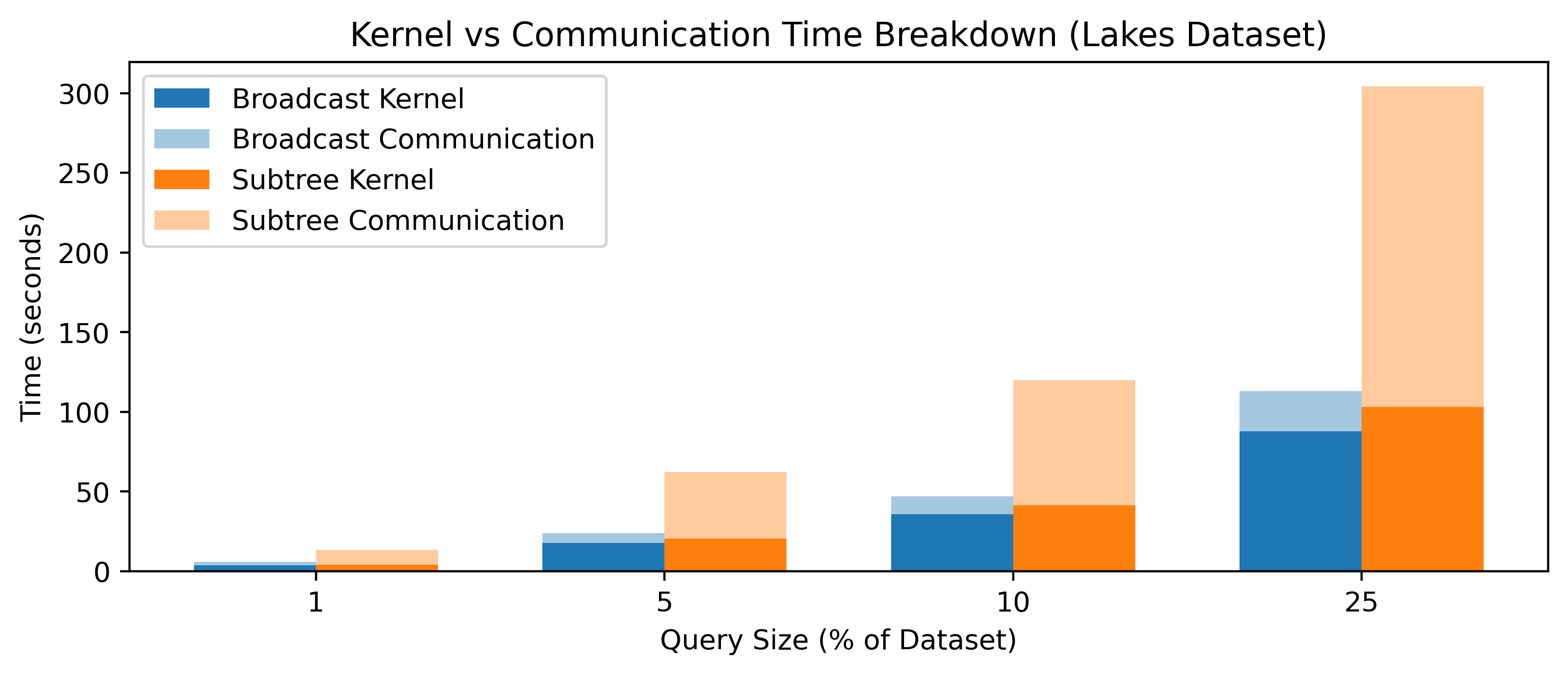}
  \caption{Kernel and communication time breakdown for the Lakes dataset (8.4M rectangles) for both Broadcast-based method and Subtree-based method on \platformPIM{}.}
  \label{fig:lakes_comm_breakdown}
\end{figure}

Figure \ref{fig:lakes_comm_breakdown} explains this behavior by breaking down kernel and communication time for the Lakes dataset. We observe similar performance on the other datasets, so we omit those results. The subtree-based method is clearly communication dominated: communication time exceeds kernel execution time at all query sizes and grows rapidly with increasing query volume due to repeated subtree transfers and per-DPU data movement. In contrast, the broadcast-based approach significantly reduces communication overhead by broadcasting the upper R-tree levels once and reusing them across all queries, resulting in a more balanced execution profile where kernel time becomes the dominant cost. Together, these results demonstrate that the broadcast-based design effectively mitigates the communication bottleneck inherent in subtree-based PIM R-trees, enabling better scalability and higher end-to-end performance.


The observed performance trends vary across datasets due to differences in dataset size and memory behavior. The Sports dataset exhibits limited benefit from PIM execution because its smaller working set provides insufficient parallel work to amortize host–DPU overheads. The Lakes dataset represents an intermediate case, where larger data volume and reduced cache effectiveness allow the broadcast-based design to mitigate communication costs and achieve consistent end-to-end speedups. In contrast, the Synthetic dataset exhibits highly irregular and memory-dominated access patterns, maximizing DPU utilization and allowing PIM execution to achieve the largest performance gains. Overall, these results show that minimizing communication is critical for scalable PIM-based spatial query processing.

\begin{table*}[t]
\centering
 \footnotesize 
\caption{Performance comparison of \approachBroadcastPIM and \approachPIMBaseline entirely run on \platformPIM{} with DPU: 2540, tasklets: 11.}
\label{tab:broadcast_vs_subtree}
\begin{tabular}{|l|r|r|r|r|r|r|r|r|r|r|r|}
\hline
\multirow{2}{*}{Dataset} &
\multirow{2}{*}{Query Size} &
\multicolumn{5}{c|}{\textbf{\approachBroadcastPIM}} &
\multicolumn{5}{c|}{\textbf{\approachPIMBaseline}} \\
\cline{3-12}
& &
\makecell{Host\\(s)} &
\makecell{Kernel\\(s)} &
\makecell{Total\\DPU}  &
\makecell{Kernel\\Speedup} &
\makecell{End-to-End\\Speedup} &
\makecell{Host\\(s)} &
\makecell{Kernel\\(s)} &
\makecell{Total\\DPU} &
\makecell{Kernel\\Speedup} &
\makecell{End-to-End\\Speedup} \\
\hline

\multirow{4}{*}{Sports}
& 9,998   & 0.41 & 0.30 & 1.10 & 1.35 & 0.37 & 0.27 & 0.33 & 2.25 & 0.82 & 0.12 \\
& 49,986  & 2.00 & 1.50 & 2.77 & 1.34 & 0.72 & 1.31 & 1.64 & 7.20 & 0.80 & 0.18 \\
& 99,971  & 3.99 & 3.00 & 4.85 & 1.33 & 0.82 & 2.62 & 3.28 & 13.38 & 0.80 & 0.20 \\
& 249,928 & 9.95 & 7.52 & 11.07 & 1.32 & 0.90 & 6.57 & 8.21 & 32.02 & 0.80 & 0.21 \\
\hline

\multirow{4}{*}{Lakes}
& 84,194   & 12.95 & 3.61 & 6.00 & 3.59 & 2.16 & 5.93 & 4.13 & 13.22 & 1.44 & 0.45 \\
& 420,967  & 64.35 & 17.57 & 23.85 & 3.66 & 2.70 & 29.64 & 20.61 & 62.38 & 1.44 & 0.48 \\
& 841,932  & 130.12 & 35.92 & 46.93 & 3.62 & 2.77 & 58.89 & 41.24 & 119.88 & 1.43 & 0.49 \\
& 2,104,829& 319.57 & 87.72 & 112.90 & 3.64 & 2.83 & 148.98 & 103.19 & 304.23 & 1.44 & 0.49 \\
\hline

\multirow{4}{*}{Synthetic}
& 159,328   & 23.52 & 1.55 & 5.31 & 15.17 & 4.43 & 17.23 & 0.65 & 18.20 & 26.51 & 0.95 \\
& 798,037   & 117.75 & 7.76 & 17.49 & 15.18 & 6.73 & 86.34 & 3.25 & 90.78 & 26.57 & 0.95 \\
& 1,597,493 & 236.19 & 15.54 & 32.78 & 15.20 & 7.21 & 173.81 & 6.54 & 167.20 & 26.58 & 1.04 \\
& 3,998,921 & 594.22 & 39.03 & 79.26 & \textbf{15.23} & \textbf{7.50} & 433.56 & 16.33 & 395.04 & 26.55 & 1.10 \\
\hline

\end{tabular}
\end{table*}

\subsection{Strong Scaling}
\label{sec:scaling}

\begin{figure}[t]
    \centering
    \includegraphics[width=\linewidth]{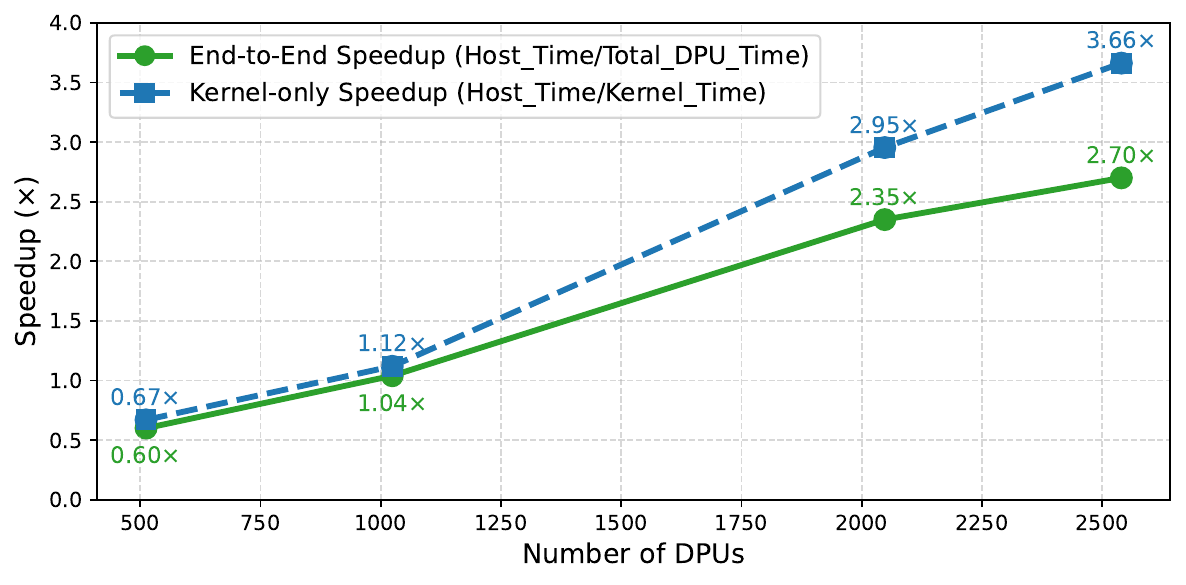}
    \caption{Strong scaling of broadcast based PIM R-tree search on fixed dataset  (8.4M rectangles). Kernel-only and end-to-end speedups improve significantly as more DPUs participate.}
    \label{fig:strong-scaling}
\end{figure}

To evaluate scalability with fixed problem size, we keep both the dataset (8.4M rectangles) and the query set (420{,}967 rectangles) fixed, while increasing the number of DPUs from 512 to 2{,}540. Figure~\ref{fig:strong-scaling} shows that both kernel-only (\PIMHost/\PIMKernel) and end-to-end speedups (\PIMHost/\PIMEtwoE) improve steadily as the number of DPUs increases, demonstrating effective strong scaling of the broadcast-based PIM R-tree. Kernel speedup grows more rapidly than end-to-end speedup, reflecting the increasing impact of fixed host to DPU communication overheads at higher DPU counts, while overall results indicate that the proposed design effectively exploits additional DPU parallelism for a fixed problem size.

\begin{figure}[t]
    \centering
    \includegraphics[width=\linewidth]{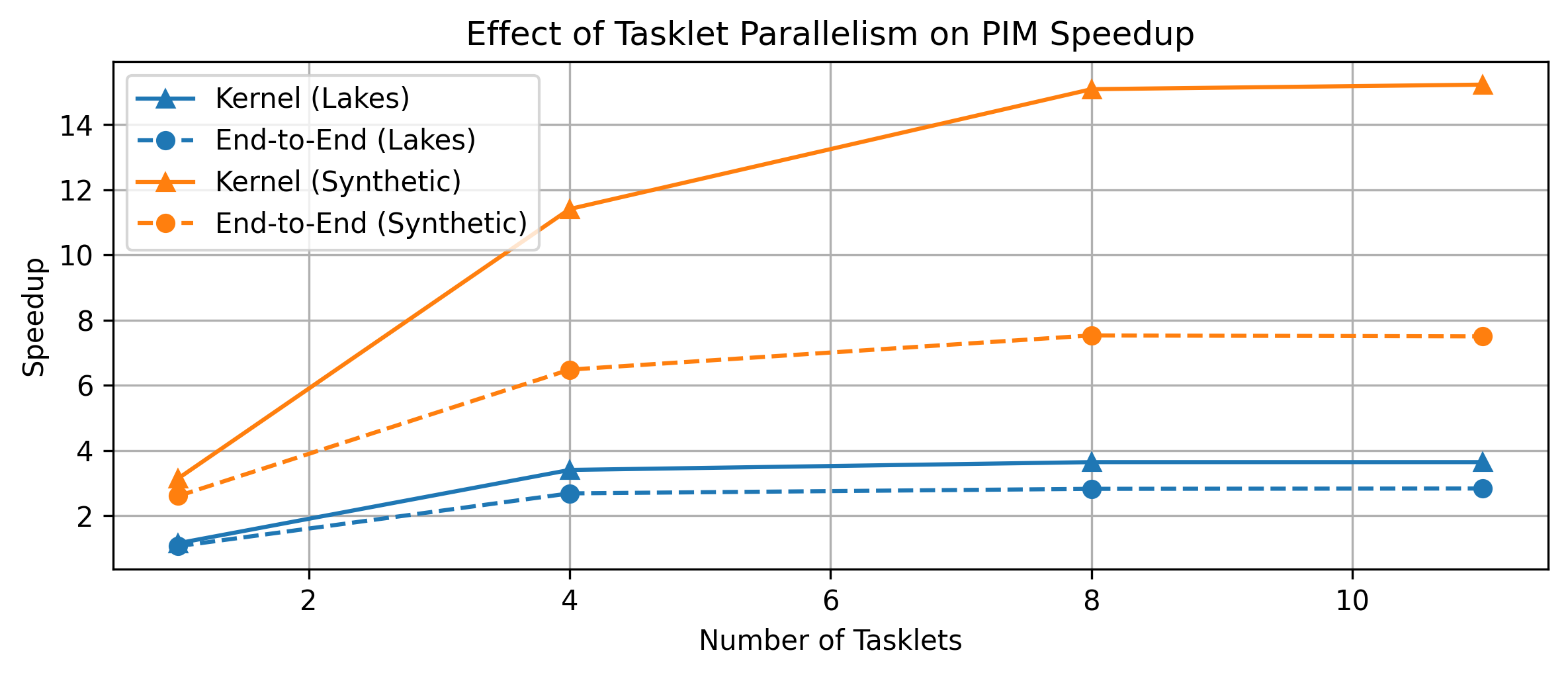}
\caption{Effect of tasklet parallelism on PIM performance. Kernel and end-to-end speedups are shown for the Lakes and Synthetic datasets as the number of tasklets per DPU increases. }
    \label{fig:tasklet}
\end{figure}

In addition to scaling across DPUs, we evaluate intra-DPU strong scaling by varying the number of tasklets per DPU. Figure \ref{fig:tasklet} shows kernel and end-to-end speedup as a function of tasklet count for the Lakes and Synthetic datasets. Speedup saturates beyond 8--11 tasklets, indicating MRAM bandwidth contention limits intra-DPU scalability. 

\subsection{Batchwise Timing Breakdown}

\begin{figure}[t]
    \centering
    \includegraphics[width=\linewidth]{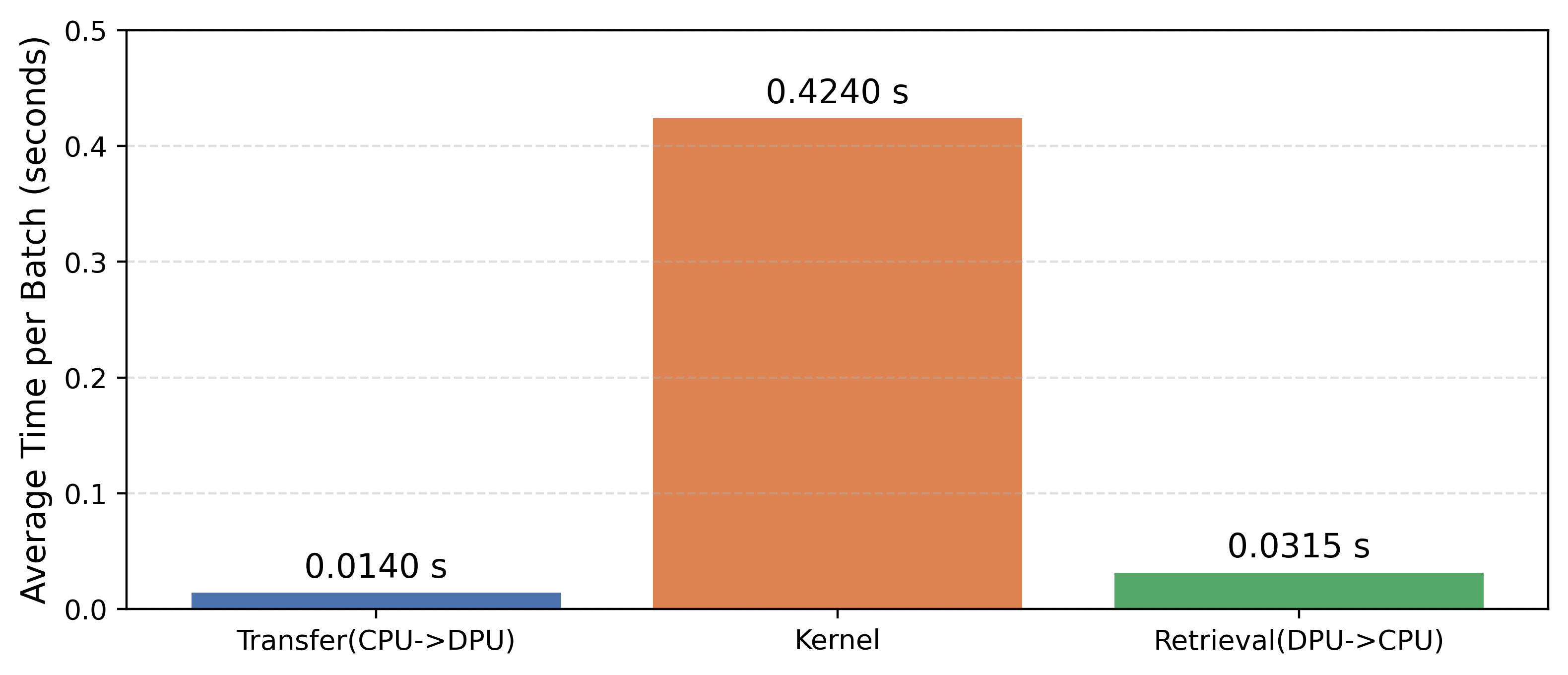}
    \caption{Average per-batch timing: host-to-DPU transfer, DPU kernel execution, and result retrieval of Lake dataset in \platformPIM{}.}
    \label{fig:batchwise}
\end{figure}

To understand the runtime behavior of the PIM execution pipeline, we measure the per-batch latency of $(i)$ query transfer,  $(ii)$ kernel execution time from cycle-accurate DPU counters, and, $(iii)$ result retrieval. Figure~\ref{fig:batchwise} shows the average time of these steps across all batches for Lakes dataset. It clearly shows that for our proposed Broadcast-based method, communication is not dominant. 

\subsection{Memory-Centric Performance Analysis}

R-tree range query processing on PIM is dominated by memory access rather than arithmetic computation. To characterize this behavior, we collect lightweight DPU-side counters that track node visits, rectangle tests, and MRAM traffic during kernel execution. These measurements provide insight into the memory access patterns that largely determine DPU kernel performance. During execution, each DPU traverses its assigned leaf slice using bounding-box filtering followed by per-rectangle intersection tests. As a result, MRAM traffic is primarily generated by reading node metadata and streaming leaf rectangles, while MRAM writes are limited to per-query result counts. Table~\ref{tab:mram-profile} summarizes the aggregate memory-access statistics for the largest real workload (8.4M rectangles, 420{,}967 queries, 2{,}540 DPUs). Across all DPUs, the kernel reads approximately 539~GB from MRAM and writes 8~GB, with more than 5.2~billion rectangle intersection tests performed.

\begin{table}[t]
\centering
\caption{Aggregate MRAM access profile of the DPU kernel.}
\label{tab:mram-profile}
\begin{tabular}{l r}
\toprule
\textbf{Metric} & \textbf{Value} \\
\midrule
Total MRAM bytes read        & 538{,}851~MB \\
Total MRAM bytes written     & 8{,}157~MB \\
Total MRAM traffic           & 547{,}009~MB \\
Nodes visited                & 19.3~billion \\
Rectangles tested            & 5.28~billion \\
Kernel execution time        & 23.48~s \\
\midrule
Attained aggregate MRAM bandwidth & 24.4~GB/s \\
\bottomrule
\end{tabular}
\end{table}
The measured aggregate bandwidth of 24.4~GB/s indicates that DPU kernel execution time closely tracks the volume of MRAM accesses rather than compute throughput. Although this bandwidth does not saturate the peak capacity of the UPMEM system, it reflects the memory-dominated nature of R-tree traversal, where each rectangle test performs only a small number of integer comparisons. As subtree size and rectangle density increase, kernel runtime scales proportionally with MRAM access volume, while host to DPU communication remains a secondary contributor to end-to-end time. These results indicate that PIM execution benefits spatial range queries primarily by minimizing data movement through near-data execution within DRAM, rather than by saturating the peak memory bandwidth of the UPMEM system.


\subsection{Energy Efficiency Analysis}
\label{sec:energy}

We evaluate energy efficiency by measuring \emph{active} system power and
execution time separately for the CPU search phase and the DPU kernel phase in \platformPIM{}. Background power states are measured only to characterize baseline system consumption and are not included in the energy calculations. Specifically, the system draws 14.5~W in standby (server off), approximately 433~W in
idle (no application running), and 528--530~W in
interactive idle (SSH session active).

For CPU-only execution, energy is computed using the active power measured while the CPU performs overlap checking, multiplied by the CPU search time. For PIM execution, energy is computed using the active power measured during DPU kernel execution (i.e., while query batches are executing on the DPUs), multiplied by the kernel execution time. During the CPU phase, active system power ranges from 567--571~W, whereas during DPU kernel execution it increases modestly to 590--601~W. Although DPU execution draws slightly higher instantaneous power, its substantially shorter kernel runtime results in significantly lower total energy consumption.

\begin{table}[t]
\centering
\footnotesize
\setlength{\tabcolsep}{4pt}
\renewcommand{\arraystretch}{1.15}
\caption{Energy comparison between CPU search phase and the DPU kernel phase in \platformPIM{}. Energy efficiency is computed as the ratio of CPU energy to DPU energy. The same execution phases and values are also reported in Table~\ref{tab:broadcast_vs_subtree}.}

\label{tab:energy}
\begin{tabular}{llrrrrr}
\toprule
Dataset &
\multicolumn{1}{c}{Query} &
\multicolumn{2}{c}{Runtime (s)} &
\multicolumn{2}{c}{Energy (kJ)} &
Energy \\
\cline{3-6}
& \multicolumn{1}{c}{Size} &
CPU & DPU &
CPU & DPU &
Efficiency \\
\midrule

\multirow{4}{*}{Sports }
& 1\%  & 0.41 & 0.30 & 0.23 & 0.18 & 1.28 \\
& 5\%  & 2.00 & 1.50 & 1.14 & 0.89 & 1.28 \\
& 10\% & 3.99 & 3.00 & 2.27 & 1.79 & 1.27 \\
& 25\% & 9.95 & 7.52 & 5.66 & 4.48 & 1.26 \\
\midrule

\multirow{4}{*}{Lakes}
& 1\%  & 12.95 & 3.61  & 7.37  & 2.15  & 3.43 \\
& 5\%  & 64.35 & 17.57 & 36.62 & 10.47 & 3.50 \\
& 10\% & 130.12 & 35.92 & 74.04 & 21.41 & 3.46 \\
& 25\% & 319.57 & 87.72 & 179.36 & 51.76 & 3.51 \\
\midrule

\multirow{4}{*}{Synthetic}
& 1\%  & 23.52 & 1.55 & 13.38 & 0.92 & 14.54 \\
& 5\%  & 117.75 & 7.76 & 67.00 & 4.62 & 14.50 \\
& 10\% & 236.19 & 15.54 & 134.39 & 9.26 & 14.51 \\
& 25\% & 594.22 & 39.03 & 338.11 & 23.26 & 14.54 \\
\bottomrule
\end{tabular}
\end{table}

Table~\ref{tab:energy} reports runtime and energy consumption for CPU-only and PIM-accelerated execution across three datasets. For the Sports dataset, energy savings are modest (about 1.26--1.28$\times$), as the dataset fits well in CPU caches and CPU execution already benefits from high locality. In contrast, for the Lakes dataset, PIM execution consistently achieves a stable energy reduction of approximately 3.4--3.5$\times$ across all query sizes. The Synthetic dataset shows the largest gains, exceeding 14$\times$ energy efficiency across all query sizes. This reflects its highly memory-intensive access pattern, where CPU execution suffers from poor cache reuse, while PIM benefits from near-data processing and parallel MRAM access. Overall, Table~\ref{tab:energy} demonstrates that energy efficiency improves with dataset size and memory intensity, highlighting the advantage of PIM architectures for large-scale, memory-bound spatial workloads.

\section{Conclusion and Future Work}

This work demonstrates the first implementation of hierarchical R-tree range queries onto a commercial PIM system. We introduce a broadcast-based execution strategy that exploits the filtering semantics of R-trees by broadcasting upper levels to all DPUs and distributing lower-level subtrees for parallel batched execution. Compared to a subtree-based PIM baseline, this design avoids per-DPU subtree transfers and prevents host–DPU communication from dominating execution, enabling scalable query processing on UPMEM hardware.

Experimental results on real and synthetic datasets using up to 2,540 DPUs show that the proposed approach delivers consistent kernel and end-to-end speedups over an optimized CPU R-tree baseline while significantly improving energy efficiency. In the Lakes dataset, strong scaling reduces kernel time by up to 3.66× and achieves a 2.70× end-to-end speedup, while consuming approximately 3.4× less energy than CPU execution in the PIM server. These results highlight the importance of communication-aware index design and execution to enable efficient and energy-efficient spatial query processing on PIM architectures.

While broadcast-based execution is most effective for large, memory-intensive workloads, performance gains are limited for small or cache-efficient datasets where communication overheads cannot be fully amortized. Based on this observation, we identify two promising directions for future work. First, we plan to explore the use of multiple DPU sets to overlap computation with data transfers, which is especially important for workloads with limited per-query computation. Second, we aim to develop an analytical performance model for PIM-based spatial queries that captures computation, communication, and memory-access costs, enabling principled decisions about when PIM execution is preferable to multi-core CPU processing.


\section*{Acknowledgments}
This material is based on work supported by the National Science Foundation under Grants No.~2344578, No.~2402987 and No.~2402988.

\bibliographystyle{ieeetr}
\bibliography{references}

@inproceedings{prasad2015gpu,
  title={{GPU-based Parallel R-tree Construction and Querying}},
  author={Prasad, Sushil K and McDermott, Michael and He, Xi and Puri, Satish},
  booktitle={2015 IEEE International Parallel and Distributed Processing Symposium Workshop},
  pages={618--627},
  year={2015},
  organization={IEEE}
}

@inproceedings{puri2013mapreduce,
  title={{MapReduce algorithms for GIS polygonal overlay processing}},
  author={Puri, Satish and Agarwal, Dinesh and He, Xi and Prasad, Sushil K},
  booktitle={2013 IEEE International Symposium on Parallel \& Distributed Processing, Workshops and Phd Forum},
  pages={1009--1016},
  year={2013},
  organization={IEEE}
}

@INPROCEEDINGS{rtreeGPU2011,
  author={Yu, Boseon and Kim, Hyunduk and Choi, Wonik and Kwon, Dongseop},
  booktitle={2011 IEEE Ninth International Conference on Dependable, Autonomic and Secure Computing}, 
  title={{Parallel Range Query Processing on R-Tree with Graphics Processing Unit}}, 
  year={2011},
  volume={},
  number={},
  pages={1235-1242},
  doi={10.1109/DASC.2011.200}
}

@inproceedings{rtree2000design,
  title={{A Design of Parallel R-tree on Cluster of Workstations}},
  author={Shuhua, Lai and Fenghua, Zhu and Yongqiang, Sun},
  booktitle={International Workshop on Databases in Networked Information Systems},
  pages={119--133},
  year={2000},
  organization={Springer}
}

@article{rtreeRDMA,
author = {Xiao, Mengbai and Wang, Hao and Geng, Liang and Lee, Rubao and Zhang, Xiaodong},
title = {{An RDMA-enabled In-memory Computing Platform for R-tree on Clusters}},
year = {2022},
issue_date = {June 2022},
publisher = {Association for Computing Machinery},
address = {New York, NY, USA},
volume = {8},
number = {2},
issn = {2374-0353},
url = {https://doi.org/10.1145/3503513},
doi = {10.1145/3503513},
journal = {ACM Trans. Spatial Algorithms Syst.},
month = feb,
articleno = {15},
numpages = {26}
}

@INPROCEEDINGS{catFishRDMA,
  author={Xiao, Mengbai and Wang, Hao and Geng, Liang and Lee, Rubao and Zhang, Xiaodong},
  booktitle={2019 IEEE 39th International Conference on Distributed Computing Systems (ICDCS)}, 
  title={{Catfish: Adaptive RDMA-enabled R-Tree for Low Latency and High Throughput}}, 
  year={2019},
  volume={},
  number={},
  pages={164-175},
  doi={10.1109/ICDCS.2019.00025}
}

@inproceedings{libRTS,
author = {Geng, Liang and Lee, Rubao and Zhang, Xiaodong},
title = {{LibRTS: A Spatial Indexing Library by Ray Tracing}},
year = {2025},
isbn = {9798400714436},
publisher = {Association for Computing Machinery},
address = {New York, NY, USA},
url = {https://doi.org/10.1145/3710848.3710850},
doi = {10.1145/3710848.3710850},
booktitle = {Proceedings of the 30th ACM SIGPLAN Annual Symposium on Principles and Practice of Parallel Programming},
pages = {396–411},
numpages = {16},
location = {Las Vegas, NV, USA},
series = {PPoPP '25}
}

@inproceedings{rtreeSpatialJoin22,
author = {Yang, Jie and Puri, Satish and Zhou, Hui},
title = {Fine-grained dynamic load balancing in spatial join by work stealing on distributed memory},
year = {2022},
isbn = {9781450395298},
publisher = {Association for Computing Machinery},
address = {New York, NY, USA},
url = {https://doi.org/10.1145/3557915.3560936},
doi = {10.1145/3557915.3560936},
booktitle = {Proceedings of the 30th International Conference on Advances in Geographic Information Systems},
articleno = {2},
numpages = {12},
location = {Seattle, Washington},
series = {SIGSPATIAL '22}
}

@INPROCEEDINGS{hifire19,
  author={Liu, Yiming and Yang, Jie and Puri, Satish},
  booktitle={2019 IEEE 26th International Conference on High Performance Computing, Data, and Analytics (HiPC)}, 
  title={{Hierarchical Filter and Refinement System Over Large Polygonal Datasets on CPU-GPU}}, 
  year={2019},
  volume={},
  number={},
  pages={141-151},
  doi={10.1109/HiPC.2019.00027}
}

@inproceedings{spjoin16,
author = {Aghajarian, Danial and Puri, Satish and Prasad, Sushil},
title = {{GCMF: an efficient end-to-end spatial join system over large polygonal datasets on GPGPU platform}},
year = {2016},
isbn = {9781450345897},
publisher = {Association for Computing Machinery},
address = {New York, NY, USA},
url = {https://doi.org/10.1145/2996913.2996982},
doi = {10.1145/2996913.2996982},
booktitle = {Proceedings of the 24th ACM SIGSPATIAL International Conference on Advances in Geographic Information Systems},
articleno = {18},
numpages = {10},
location = {Burlingame, California},
series = {SIGSPATIAL '16}
}

@INPROCEEDINGS{rtreeOverlay,
  author={Puri, Satish and Prasad, Sushil K.},
  booktitle={2015 15th IEEE/ACM International Symposium on Cluster, Cloud and Grid Computing}, 
  title={{A Parallel Algorithm for Clipping Polygons with Improved Bounds and a Distributed Overlay Processing System Using MPI}}, 
  year={2015},
  volume={},
  number={},
  pages={576-585},
  doi={10.1109/CCGrid.2015.43}
}

@inproceedings{you2013parallel,
  title={{Parallel spatial query processing on gpus using R-trees}},
  author={You, Simin and Zhang, Jianting and Gruenwald, Le},
  booktitle={Proceedings of the 2nd ACM SIGSPATIAL international workshop on analytics for big geospatial data},
  pages={23--31},
  year={2013}
}

@article{kim2013parallel,
  title={{Parallel multi-dimensional range query processing with R-trees on GPU}},
  author={Kim, Jinwoong and Kim, Sul-Gi and Nam, Beomseok},
  journal={Journal of Parallel and Distributed Computing},
  volume={73},
  number={8},
  pages={1195--1207},
  year={2013},
  publisher={Elsevier}
}

@inproceedings{schnitzer1999master,
  title={{Master-client R-trees: A new parallel R-tree architecture}},
  author={Schnitzer, Bernd and Leutenegger, Scott T},
  booktitle={Proceedings. Eleventh International Conference on Scientific and Statistical Database Management},
  pages={68--77},
  year={1999},
  organization={IEEE}
}

@article{kamel1992parallel,
  title={{Parallel R-trees}},
  author={Kamel, Ibrahim and Faloutsos, Christos},
  journal={ACM SIGMOD Record},
  volume={21},
  number={2},
  pages={195--204},
  year={1992},
  publisher={ACM New York, NY, USA}
}

@article{ghosh2019ucr,
  title={{UCR-STAR: The UCR spatio-temporal active repository}},
  author={Ghosh, Saheli and Vu, Tin and Eskandari, Mehrad Amin and Eldawy, Ahmed},
  journal={SIGSPATIAL Special},
  volume={11},
  number={2},
  pages={34--40},
  year={2019},
  publisher={ACM New York, NY, USA}
}

@misc{upmem2018whitepaper,
  title        = {{Introduction to UPMEM PIM: Processing-in-Memory (PIM) on DRAM Accelerator (White Paper)}},
  author       = {{UPMEM}},
  year         = {2018},
  howpublished = {White paper, UPMEM, Grenoble, France}
}

@inproceedings{devaux2019true,
  author       = {Devaux, Fabrice},
  title        = {{The true processing in memory accelerator}},
  booktitle    = {Proceedings of the IEEE Hot Chips 31 Symposium (HCS)},
  year         = {2019},
  pages        = {1--24},
  organization = {IEEE Computer Society}
}

@article{gomez2022benchmarking,
  title={Benchmarking a new paradigm: Experimental analysis and characterization of a real processing-in-memory system},
  author={G{\'o}mez-Luna, Juan and El Hajj, Izzat and Fernandez, Ivan and Giannoula, Christina and Oliveira, Geraldo F and Mutlu, Onur},
  journal={IEEE Access},
  volume={10},
  pages={52565--52608},
  year={2022},
  publisher={IEEE}
}

@misc{spiderweb2021,
  author = {Pulona Katiyar and Tian Wu and Sara Migliorini and Alberto Belussi and Ahmed Eldawy},
  title = {SpiderWeb: A Spatial Data Generator for Big Spatial Data Applications},
  year = {2021},
  howpublished = {\url{https://spider.cs.ucr.edu/}}
}

@inproceedings{kang2023pim,
  title={PIM-trie: A Skew-resistant Trie for Processing-in-Memory},
  author={Kang, Hongbo and Zhao, Yiwei and Blelloch, Guy E and Dhulipala, Laxman and Gu, Yan and McGuffey, Charles and Gibbons, Phillip B},
  booktitle={Proceedings of the 35th ACM Symposium on Parallelism in Algorithms and Architectures},
  pages={1--14},
  year={2023}
}

@inproceedings{leutenegger1997str,
  title={{STR: A simple and efficient algorithm for R-tree packing}},
  author={Leutenegger, Scott T and Lopez, Mario A and Edgington, Jeffrey},
  booktitle={Proceedings 13th international conference on data engineering},
  pages={497--506},
  year={1997},
  organization={IEEE}
}

@article{cai2024pimpam,
  title={Pimpam: Efficient graph pattern matching on real processing-in-memory hardware},
  author={Cai, Shuangyu and Tian, Boyu and Zhang, Huanchen and Gao, Mingyu},
  journal={Proceedings of the ACM on Management of Data},
  volume={2},
  number={3},
  pages={1--25},
  year={2024},
  publisher={ACM New York, NY, USA}
}

@article{baumstark2023accelerating,
  title={Accelerating large table scan using processing-in-memory technology},
  author={Baumstark, Alexander and Jibril, Muhammad Attahir and Sattler, Kai-Uwe},
  journal={Datenbank-Spektrum},
  volume={23},
  number={3},
  pages={199--209},
  year={2023},
  publisher={Springer}
}

@article{lim2023design,
  title={Design and analysis of a processing-in-dimm join algorithm: A case study with upmem dimms},
  author={Lim, Chaemin and Lee, Suhyun and Choi, Jinwoo and Lee, Jounghoo and Park, Seongyeon and Kim, Hanjun and Lee, Jinho and Kim, Youngsok},
  journal={Proceedings of the ACM on Management of Data},
  volume={1},
  number={2},
  pages={1--27},
  year={2023},
  publisher={ACM New York, NY, USA}
}

@inproceedings{guttman1984r,
  title={R-trees: A dynamic index structure for spatial searching},
  author={Guttman, Antonin},
  booktitle={Proceedings of the 1984 ACM SIGMOD international conference on Management of data},
  pages={47--57},
  year={1984}
}

@article{wulf1995hitting,
  title={Hitting the memory wall: Implications of the obvious},
  author={Wulf, Wm A and McKee, Sally A},
  journal={ACM SIGARCH computer architecture news},
  volume={23},
  number={1},
  pages={20--24},
  year={1995},
  publisher={ACM New York, NY, USA}
}

@article{patterson2002case,
  title={{A case for intelligent RAM}},
  author={Patterson, David and Anderson, Thomas and Cardwell, Neal and Fromm, Richard and Keeton, Kimberly and Kozyrakis, Christoforos and Thomas, Randi and Yelick, Katherine},
  journal={IEEE micro},
  volume={17},
  number={2},
  pages={34--44},
  year={2002},
  publisher={IEEE}
}

@article{ahn2015pim,
  title={{PIM-enabled instructions: A low-overhead, locality-aware processing-in-memory architecture}},
  author={Ahn, Junwhan and Yoo, Sungjoo and Mutlu, Onur and Choi, Kiyoung},
  journal={ACM SIGARCH Computer Architecture News},
  volume={43},
  number={3S},
  pages={336--348},
  year={2015},
  publisher={ACM New York, NY, USA}
}

@article{balasubramonian2014near,
  title={Near-data processing: Insights from a micro-46 workshop},
  author={Balasubramonian, Rajeev and Chang, Jichuan and Manning, Troy and Moreno, Jaime H and Murphy, Richard and Nair, Ravi and Swanson, Steven},
  journal={IEEE Micro},
  volume={34},
  number={4},
  pages={36--42},
  year={2014},
  publisher={IEEE}
}

@inproceedings{seshadri2017ambit,
  title={{Ambit: In-memory accelerator for bulk bitwise operations using commodity DRAM technology}},
  author={Seshadri, Vivek and Lee, Donghyuk and Mullins, Thomas and Hassan, Hasan and Boroumand, Amirali and Kim, Jeremie and Kozuch, Michael A and Mutlu, Onur and Gibbons, Phillip B and Mowry, Todd C},
  booktitle={Proceedings of the 50th Annual IEEE/ACM International Symposium on Microarchitecture},
  pages={273--287},
  year={2017}
}

@inproceedings{zhao2026pim,
  title={PIM-zd-tree: A Fast Space-Partitioning Index Leveraging Processing-in-Memory},
  author={Zhao, Yiwei and Kang, Hongbo and Men, Ziyang and Gu, Yan and Blelloch, Guy E and Dhulipala, Laxman and McGuffey, Charles and Gibbons, Phillip B},
  booktitle={Proceedings of the 31st ACM SIGPLAN Annual Symposium on Principles and Practice of Parallel Programming},
  pages={480--495},
  year={2026}
}

@inproceedings{friesel2023full,
  title={A full-system perspective on upmem performance},
  author={Friesel, Birte and L{\"u}tke Dreimann, Marcel and Spinczyk, Olaf},
  booktitle={Proceedings of the 1st Workshop on Disruptive Memory Systems},
  pages={1--7},
  year={2023}
}

@article{kim2025no,
  title={No Cap, This Memory Slaps: Breaking Through the Memory Wall of Transactional Database Systems with Processing-in-Memory},
  author={Kim, Hyoungjoo and Zhao, Yiwei and Pavlo, Andrew and Gibbons, Phillip B},
  journal={Proceedings of the VLDB Endowment},
  volume={18},
  number={11},
  pages={4241--4254},
  year={2025},
  publisher={VLDB Endowment}
}
\end{document}